\documentclass[journal=jacsat,manuscript=article]{achemso}
\setkeys{acs}{doi = true}
\SectionNumbersOn

\usepackage{graphicx}
\usepackage{dcolumn}
\usepackage{bm}
\usepackage{siunitx}
\usepackage[utf8]{inputenc}
\usepackage[T1]{fontenc}
\usepackage{threeparttable}
\usepackage{chemfig}
\usepackage{mathptmx}
\usepackage{etoolbox}
\usepackage{amsmath}
\usepackage{caption}
\usepackage{subcaption} 
\usepackage{url} 
\usepackage{sidecap} 
\usepackage{doi}

\newcommand{\ionp}{CCH$^+$ }
\newcommand{\ion}{CCH$^+$}
\newcommand{\wn}{cm$^{-1}$}
\newcommand{\wns}{cm$^{-1}$ }

\newcommand{\felix}{HFML-FELIX, Toernooiveld 7, 6525 ED Nijmegen, The Netherlands}
\newcommand{\imm}{Institute for Molecules and Materials, Radboud University, Heyendaalseweg 135, 6525 AJ Nijmegen, The Netherlands}
\newcommand{\desy}{Photon Science Division, Deutsches-Elektronen-Synchrotron DESY, Notkestr. 85, 22607 Hamburg, Germany}
\newcommand{\cologne}{I. Physikalisches Institut, Universit\"{a}t zu K\"{o}ln, Zülpicher Str. 77, 50937 K\"{o}ln, Germany}

\title{High-resolution ro-vibrational and rotational spectroscopy of the open-shell, linear \ionp ion ($^3\Pi$)}

\author{Kim Steenbakkers}
\affiliation{\felix}
\alsoaffiliation{\imm}

\author{Weslley G. D. P. Silva}
\affiliation{\cologne}

\author{Oskar Asvany}
\affiliation{\cologne}

\author{Gerrit C. Groenenboom}
\affiliation{Theoretical Chemistry, Institute for Molecules and Materials, Heyendaalseweg 135, 6525 AJ Nijmegen, The Netherlands}

\author{Pavol Jusko}
\affiliation{Max Planck Institute for Extraterrestrial Physics, Gießenbachstraße 1, 
85748 Garching, Germany}

\author{Britta Redlich}
\affiliation{\felix}
\alsoaffiliation{\imm}
\alsoaffiliation{\desy}

\author{Sandra Br\"{u}nken}
\affiliation{\felix}
\alsoaffiliation{\imm}
\email{sandra.bruenken@ru.nl}

\author{Stephan Schlemmer}
\affiliation{\cologne}
\email{schlemmer@ph1.uni-koeln.de}

\date{\today}

\begin{document}

\begin{abstract}
 
In this work, we report on the high-resolution infrared spectrum of \ionp ($^3\Pi$) recorded in the range $3066-3184$~\wns by means of leak-out spectroscopy. This spectral range covers the fundamental of the CH stretching mode and a highly excited bending vibrational mode. Based on this data (385 ro-vibrational lines), accurate spectroscopic descriptions of the ground and the two vibrationally excited states of \ionp were obtained. Besides the band origins, spin-orbit coupling constants,  rotational constants, centrifugal distortion constants and $\Lambda$-doubling constants for the ground and excited vibrational states have been derived. This effective Hamiltonian analysis allowed a search for pure rotational lines of \ionp in its electronic and vibrational ground state using a two-color millimeterwave - infrared  scheme. We observed all rotational transitions from $J^{\prime\prime} = 2$ up to $J^{\prime\prime} = 6$ within the $\Omega = 2$ lowest energy fine structure component with resolved hyperfine splittings. This data has already guided the first detection of \ionp in space toward the Orion Bar photo-dissociation region, and has the potential to support further astronomical searches for \ionp either through radio or infrared spectroscopy, for example with the James Webb Space Telescope.

\end{abstract}

\section{Introduction}

The ethynyl radical cation, \ionp ($^3\Pi$), is of significant interest from multiple perspectives. From a fundamental spectroscopic standpoint, it is intriguing because it is one of the few molecules in a $^3\Pi$ electronic state for which experimental spectroscopic data are available, although until now not through infrared (IR) spectroscopy with rotational resolution\cite{lichten1979fine,davies1987infrared,deo2004infrared,ram2010revised,simard1988high}. Studying systems in such a $^3\Pi$ electronic state is challenging due to the need to account for numerous coupling effects, including spin-orbit coupling, $\Lambda$-doubling, spin-spin coupling, and spin-rotation interaction. Moreover, bending modes are affected by Renner-Teller coupling. Earlier theoretical works on \ionp have focused on unravelling its intricate electronic structure\cite{Hashimoto1990,Koch1990,Krishnan1981}, confirming the ground electronic configuration 3$\sigma^2$4$\sigma^2$5$\sigma^1$1$\pi^3$ and its  $^3\Pi$ ground state. The complexity of \ion, however, extends beyond these considerations as shown in the theoretical work of \citet{Mehnen2018}. Not only does \ionp exhibit a $^3\Pi$ electronic ground state, but it also presents several low-lying excited electronic states which are expected to lead to additional vibronic coupling effects. On the one hand, this makes \ionp an interesting candidate for benchmarking spectroscopic models, and on the other hand, it makes its spectroscopic parameters extremely difficult to be computed \emph{ab initio}.

Beyond its spectroscopic significance, \ionp is also of considerable astrochemical interest. 
Its neutral counterpart, CCH ($^2\Sigma^+$), is a known molecule of astrochemical importance. 
Since its first detection in space in 1974 \cite{Tucker1974}  it has been observed in several other astronomical sources \cite{Ziurys2006,Dobrijevic2016,Wilson2003,Jackson1996}.
The omnipresence of CCH in space has triggered the investigation of its cationic counterpart \ionp ($^3\Pi$), whose existence in space was assumed highly probable and which has been proposed to be a key intermediate in the formation of small hydrocarbons under astrophysical conditions. \cite{Guzman2015,cuadrado2015chemistry}. Even though the radical CCH was detected 50 years ago, \ionp has only been identified recently in the Orion Bar \cite{Jacob2025}, guided by the experimental results presented here. The reason for this relatively late detection is two-fold: \ion's extreme reactivity makes it challenging to obtain experimental spectroscopic data and its complex electronic structure hinders the prediction of accurate \emph{ab initio} ro-vibrational and rotational transition frequencies. 

Earlier experimental works related to \ionp consist of translational energy spectroscopy of CCD$^+$ of \citet{Keefe1984}, which strongly indicated that \ionp exhibits a $^3\Pi$ ground state. Furthermore, the geometrical probability density functions were determined for the bending modes of \ionp through Coulomb explosion imaging \cite{Zajfman1991}. The first IR spectroscopy experiment was carried out by \citet{Andrews1999}, who measured the spectrum of \ionp in solid Ar and Ne matrices in the range 1660-1860 \wn, covering the CC stretching fundamental. Subsequently, a photoionization study on the adiabatic ionization energy of the CCH radical was performed by \citet{Gans2017}, providing the first (low-resolution) spectroscopic data on the Renner-Teller perturbed CCH bending mode of \ion. Finally, a recent He-droplet IR spectroscopic study\cite{Feinberg2023} in the range 3050-3250 \wns revealed the first data on the CH stretching fundamental of \ion, but information on its ro-vibrational structure could not be obtained due to the interaction between \ionp and the surrounding He atoms.

Our groups have now extended the experimental work extensively by recording both the broadband vibrational (350-3450 \wn), high-resolution ro-vibrational (3066-3184 \wn) and pure rotational spectrum of \ion. The implications of these works are extensive and cannot be addressed in a single study. 
Therefore, the investigation of \ionp was divided into three parts with different goals. The initial aim of our work focused on the first detection of \ionp in space and its astrochemical implications, described in  \citet{Jacob2025}, which became only possible due to the laboratory work presented here. The recording of the broadband IR spectrum of \ionp, in combination with the present work and a three-state diabatic model, allowed to unravel the intricate electronic state mixing effects between the two Renner-Teller $^3\Pi$ states and the close lying  $^3\Sigma^-$ state, as presented in \citet{Steenbakkers2025} .

In this paper, our goal is to provide accurate spectroscopic parameters required both for the detection of \ionp in space and for benchmarking the aforementioned three-state model. To address this, we first recorded the high-resolution IR spectrum of \ionp in the range 3066-3184 \wns by means of leak-out spectroscopy (LOS) \cite{Schmid2022}, described in Sections \ref{sec:rovib} and \ref{sec:IR}. The resulting spectrum covers the $v_1$ CH stretch fundamental as well as a second vibrational mode,  tentatively assigned by \citet{Steenbakkers2025}, to a highly excited bending mode. These spectra are analyzed with a standard Hamiltonian for a $\Pi$-$\Pi$ transition including the fine structures and couplings mentioned above which are laid out in Section \ref{sec:model} on the spectroscopic model, and the results are described in Section \ref{sec:fit_param}. Based on the ground state parameters obtained from the analysis of the ro-vibrational measurements, we then recorded rotational transitions (from $J^{\prime\prime} = 2$ to $J^{\prime\prime} = 6$) of \ionp within the $\Omega=2$ ground state using a two-colour millimeter-IR scheme described in detail in \citet{Asvany2023}, and here in Sections \ref{sec:rot} and \ref{sec:Rot_meas}. The accurately determined spectroscopic parameters of this combined study aided investigations of the broadband vibrational spectrum of \ionp\cite{Steenbakkers2025}, and allowed the astronomical detection of \ionp by radio astronomy based on its rotational transitions \cite{Jacob2025}.

\section{Experimental and Theoretical Methods}

\subsection{Ro-vibrational spectroscopy}
\label{sec:rovib}
The ro-vibrational and pure rotational spectra of \ionp were measured in a 22-pole cryogenic ion trap instrument (COLTRAP), which has been previously described in detail\cite{Asvany2014}. The \ionp ions were produced in a storage ion source (SIS) \emph{via} electron impact ionization (E$_{e^-} = 30$~eV) of a 3:1 He:acetylene (C$_2$H$_2$) precursor mixture, which was introduced in the source at a pressure of about 10$^{-5}$ mbar. The \ionp ions were then extracted from the source, mass selected in a quadrupole mass filter (for \emph{m/z}~=~25), and injected into the 22-pole trap\cite{asv10}, kept at a nominal temperature of 4~K for all measurements. The trap is constantly filled with He buffer gas ($\emph{n}$ $\approx$ 10$^{13}$ cm$^{-3}$) which ensures thermalisation of the incoming ions to the cryogenic temperature. Additionally, at the beginning of each trapping cycle, a 1:3 Ne:He mixture was pulsed into the trap using a piezoelectrically actuated valve. For the leak-out spectroscopic scheme the presence of Ne in the trap is essential as this atom serves as the collision partner for this method of action spectroscopy. 
The introduction of the Ne atoms into the trap as a pulsed mixture diluted in He avoids freeze out of Ne throughout the measurements and ensures stable experimental conditions at the cryogenic temperature, as shown in recent studies.\cite{gupta23,sil23b,sil24}

Once trapped, the ro-vibrational transitions of \ionp were measured using the leak-out spectroscopy (LOS) method which has been reported in detail recently\cite{Schmid2022}. In short, LOS is based on the escape of an ion from the trap after it is vibrationally excited and undergoes inelastic collisions with a neutral atomic/molecular partner (Ne in this case). In this collision a part of the vibrational excitation energy is converted into translational energy (V-T transfer) which allows the then faster ion to leave the trap and be counted by an ion detector. The action spectroscopic signal is recorded by counting the number of ions that leak from the trap upon tuning the frequency of the radiation source. 

For ro-vibrational spectroscopy, narrow bandwidth ($ < 10^{-4}$ \wn) tunable IR radiation, generated by a continuous-wave optical parametric oscillator (cw-OPO, Toptica, model TOPO) operating in the 3 $\mu$m spectral region, was used as the light source. The beam entered the vacuum chamber through a diamond window (0.6 mm thickness, Diamond Materials GmbH), irradiated the ions for 300 ms by crossing the 22-pole trap, and exited the trap instrument via a CaF$_2$ window, after which it was absorbed by a power meter. The measured power was on the order of a few hundred mW. The frequency of the IR radiation was measured by a spectrum analyzer (Bristol instruments, Model 771A-MIR), which has an accuracy of ~0.001 \wn. 

\subsection{Rotational spectroscopy}
\label{sec:rot}
To measure pure rotational transitions, we used a double resonance vibrational-rotational spectroscopy scheme based on LOS, which was recently reported in detail\cite{Asvany2023}.
In this experiment, the trapped ions are excited by both fixed frequency IR and tunable mm-wave radiation which are overlaid outside the apparatus using an ellipsoidal mirror (\emph{f}= 43.7 mm) before entering the trap \cite{asv21d}. The mm-wave radiation was produced by a microwave synthesizer (Rohde \& Schwarz, SMF 100A) which is locked to a Rb atomic clock, driving different amplifier-multiplier chains (Virginia Diodes Inc.). 
With these setups, frequencies from 82.5~GHz up to $1.1$~THz can be produced with typical output powers ranging from 25~mW to $5\times10^{-3}$~mW depending on the frequency range. It is worth noting, however, that the mm-wave power was attenuated and lowered as much as possible to minimize power broadening effects, especially at lower frequencies. Specific details for the measurements will be given in Section \ref{sec:Rot_meas}.  

\subsection{Spectroscopic Model}
\label{sec:model}

\begin{figure*}
    \centering
    \includegraphics[width=\textwidth]{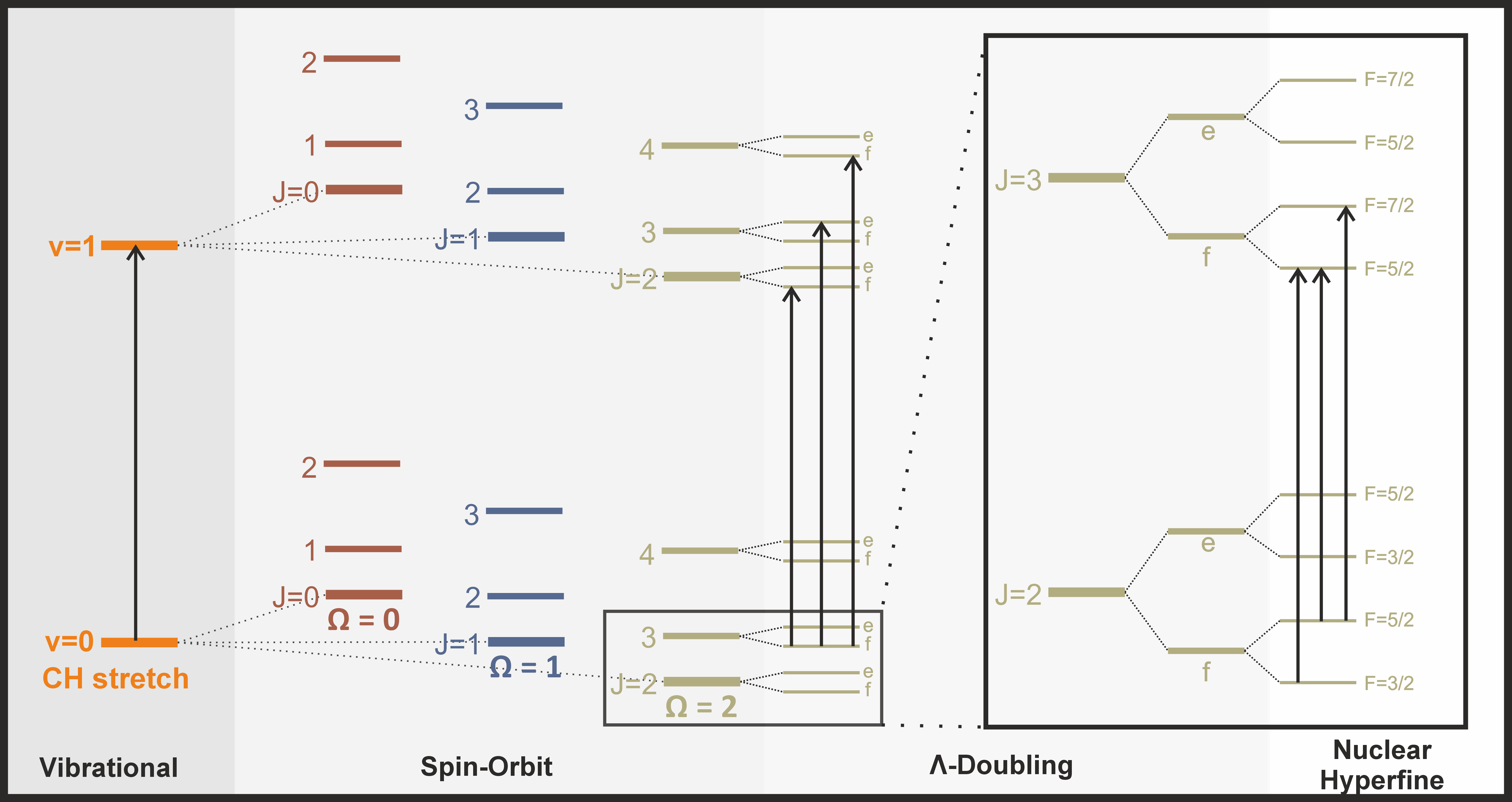}
    \caption{Energy level diagram of \ionp (not to scale), showing the splittings as a result of SO coupling, $\Lambda$-doubling and nuclear hyperfine interaction. The arrows represent exemplary transitions that are allowed following the $\Delta\Omega=0$, $\Delta J=0$,$\pm1$ and $\Delta F=0$,$\pm1$ selection rules for a vibronic $\Pi$-$\Pi$ transition.}
    \label{fig:Term_scheme}
\end{figure*}

To model the \ionp ion we must take into account its electronic, vibrational, and rotational structure, together with the coupling effects between the different angular momenta. The spectroscopic model was created within the program PGOPHER~\cite{wes17,Western2019}, where the following effective Hamiltonian was used for the ro-vibrational data: 

\begin{equation}
  \hat{H}_\mathrm{tot}= E_\mathrm{Vib} +  \hat{H}_\mathrm{SO} + \hat{H}_\mathrm{Rot} + \hat{H}_\mathrm{\Lambda} + \hat{H}_\mathrm{SS} + \hat{H}_\mathrm{SR}.
    \label{eqn:Hamil}
\end{equation}

\noindent Here the terms are ordered from large to small and the electronic energy was set to zero. After the electronic energy, vibrational energy $E_\mathrm{Vib}$, gives rise to the largest energy differences as shown on the left side of Figure \ref{fig:Term_scheme}. Here the vibrational ground state ($v=0$) and excited states ($v=1$) are displayed for the C-H stretching vibration in the grey shaded area. 

The next largest interaction is a result of the $^3\Pi$ electronic ground state of \ion. Its two unpaired electrons give rise to a total spin $S=1$ with projections on the molecular axis $\Sigma = -1, 0, 1$, which can couple to the orbital angular momentum, $L$, with its projection onto the molecular axis $\Lambda = \pm1$. This coupling leads to the formation of three non-degenerate spin-orbit (SO) states, labeled by $\Omega=\Lambda+\Sigma=0, 1, 2$, where the energy separation is given by the spin-orbit constant $A$. 
The spin-orbit coupling Hamiltonian, $\hat{H}_\mathrm{SO}$, is given by\cite{pople1960renner} 

\begin{equation}
  \hat{H}_\mathrm{SO} = A\hat{L}_z \hat{S}_z + \frac{A_{D}}{2}[\hat{R}^2,\hat{L}_z\hat{S}_z]_+ + \frac{A_{H}}{2}[\hat{R}^4,\hat{L}_z\hat{S}_z]_+,
\end{equation}

\noindent where $\hat{L}_z=\hat{\Lambda}$ and $\hat{S}_z=\hat{\Sigma}$ represent the molecule fixed components of the electronic orbital and spin angular momentum operators. The parameters $A_D$ and $A_H$ represent the centrifugal distortion constants of $A$.  

In the case of \ionp $A$ is predicted to be negative\cite{Mehnen2018}, which means that $\Omega=2$ should be the ground state, termed $F_1$, being lower in energy by $A$ than the unshifted $\Omega=\Lambda=1$ state ($F_2$) and 2$A$ lower in energy than the $\Omega=0$ state ($F_3$), as displayed (not to scale) in Figure \ref{fig:Term_scheme} for the lowest rotational states ($J$). 

The following interaction of energetic significance is the rotational motion of \ion. This specific energy ordering (i.e. rotational interaction is smaller than the spin-orbit splitting) results in \ionp to be best described by a Hund's case (a) basis, with the rotational Hamiltonian, $\hat{H}_\mathrm{Rot}$, given by\cite{hirota1985}:

\begin{equation}
 \hat{H}_\mathrm{Rot} = B\hat{R}^2-D\hat{R}^4, 
 \label{eqn:H_Rot}
\end{equation}

\noindent where $\hat{R}$ is the rotational angular momentum given by $\hat{R}=\hat{J}-\hat{L}-\hat{S}$, with $\hat{J}$ being the total angular momentum, $\hat{L}$ the electronic angular momentum and $\hat{S}$ the spin angular momentum. The rotational constant is denoted $B$, and $D$ is the centrifugal distortion of the rotation.
Based on Eq.~\ref{eqn:H_Rot} the rotational energy leads to a ladder of $J$-states increasing quadratically in $J$ as can be seen in  Figure~\ref{fig:Term_scheme} for $\Omega=2,1,0$ ($F_1, F_2, F_3$). 

Furthermore, the rotational motion of the ion can result in what is essentially an axial symmetry breaking leading to a lift of the degeneracy of the $\Lambda=-1$ and $+1$ terms. This effect is commonly referred to as $\Lambda$-doubling and causes a splitting of the rotational states. The rotationless parity of the energy levels is then given by the $e$ and $f$ labels, where $e$ refers to levels with parity $(-1)^J$ and $f$ to parity $-(-1)^J$ (see Figure \ref{fig:Term_scheme}). The effective Hamiltonian related to $\Lambda$-doubling is described by\cite{brown1979}:

\begin{align}
      \hat{H}_\mathrm{\Lambda} &= \frac{1}{2}o\{\hat{S}_+^2e^{-2i\phi}+ \hat{S}_-^2e^{+2i\phi}\} - \frac{1}{2}p\{(\hat{J}_+-\hat{S}_+)\hat{S}_+e^{-2i\phi}+ (\hat{J}_--\hat{S}_-)\hat{S}_-e^{+2i\phi}\} \nonumber\\
      &+ \frac{1}{2}q\{(\hat{J}_+-\hat{S}_+)^2e^{-2i\phi} + (\hat{J}_--\hat{S}_-)^2e^{+2i\phi}\} + \frac{1}{4}[\hat{S}_+^2e^{-2i\phi}+ \hat{S}_-^2e^{+2i\phi},o_D\hat{R}^2]_+ \nonumber\\
      &- \frac{1}{4} [(\hat{J}_+-\hat{S}_+)\hat{S}_+e^{-2i\phi}+ (\hat{J}_--\hat{S}_-)\hat{S}_-e^{+2i\phi}, p_D\hat{R}^2]_+ 
       \nonumber\\
      &+\frac{1}{4}[(\hat{J}_+-\hat{S}_+)^2e^{-2i\phi}+ (\hat{J}_--\hat{S}_-)^2e^{+2i\phi},q_D\hat{R}^2]_+,
\end{align}

\noindent where $o$, $p$, and $q$ are the $\Lambda$-doubling parameters, $o_D$, $p_D$, and $q_D$ their centrifugal distortions, and $\hat{S}_\pm$ and $\hat{R}_\pm$ the spin angular momentum and rotational angular momentum ladder operators.

Additionally, the two unpaired electrons may interact through spin-spin coupling which leads to a shift of the ro-vibrational lines and is given by\cite{hirota1985}:

\begin{equation}
      \hat{H}_\mathrm{SS} = \frac{2}{3}\lambda_{SS}(3\hat{S}_z^2-\hat{S}^2) + \frac{1}{2} \lambda_D[\frac{2}{3}(3\hat{S}_z^2-\hat{S}^2),\hat{R}^2]_+,
\end{equation}

\noindent where $\lambda_{SS}$ is the spin-spin coupling constant and $\lambda_D$ its centrifugal distortion.

Finally, the electronic spin will couple to the rotational angular momentum through spin-rotation coupling, resulting in a shift of the ro-vibrational lines as well. This Hamiltonian is described by\cite{hirota1985}:
\begin{equation}
      \hat{H}_\mathrm{SR} = \gamma(\hat{J}-\hat{S})\cdot\hat{S} + \frac{1}{2} \gamma_D[(\hat{J}-\hat{S})\cdot\hat{S},\hat{R}^2]_+ + \frac{1}{2} \gamma_H[(\hat{J}-\hat{S})\hat{S},\hat{R}^4]_+,
\end{equation}

\noindent where $\gamma$ is the spin-rotation constant and $\gamma_D$ and $\gamma_H$ its centrifugal distortion. It is worth noting that the spin-rotation coupling, described by $\gamma$, has the same effect on the energy levels as $A_D$\cite{Veseth1971}, so only one of the two parameters should be used. Here, we opted to include $A_D$, which then accounts for both effects. For the spectroscopic constants, calculated values exist only for $A$ and $B$, as reported by \citet{Mehnen2018}, which were used as initial estimates. All other parameters were determined directly from the spectrum.

The result of $\Lambda$-doubling and the other shifts of the energy levels are shown schematically in the left part of Figure~\ref{fig:Term_scheme} for $\Omega=2$ for the lowest rotational states which are all split into their respective e- and f-components. This splitting is shown enlarged in the right part of the Figure in order to visualize the hyperfine splitting which will be considered below.


On top of the electronic angular momentum interactions discussed so far, \ionp contains a single hydrogen with nuclear spin $I=1/2$, making it subjective to nuclear spin-orbit coupling \cite{hirota1985}:

\begin{equation} \hat{H}_\mathrm{IL}=a\hat{I}\cdot\hat{L}, \end{equation}

\noindent nuclear spin-electron spin interaction\cite{hirota1985}:

\begin{equation} \hat{H}_\mathrm{IS}=b_\eta\hat{I}\cdot\hat{S}, \end{equation}

\noindent and nuclear spin-electron spin dipole-dipole coupling\cite{hirota1985}:

\begin{equation} \hat{H}_\mathrm{DD}=\hat{S}\cdot{\bm T}\hat{I}, \end{equation}

\noindent where ${\bm T}$ represents a second-rank traceless tensor. These interactions are characterized by four spectroscopic parameters; $a$, $b$, $c$, and $d$, first introduced by \citet{frosch1952magnetic}.
These nuclear interactions give rise to a small splitting pattern (<30 MHz, see Figure \ref{fig:Term_scheme}), which can only be resolved in the rotational spectrum, but may be ignored in the analysis of the ro-vibrational spectrum. 

The model described here can simulate the rotational  structure of any molecule in a $^3\Pi$ electronic state 
and in a vibrational state which does not carry angular momentum. In those cases the vibronic Renner-Teller coupling effect can be ignored and each vibrational state is described by the same rotational structure model albeit with different molecular constants for each vibrational state as illustrated schematically for the $v=0$ and $v_1=1$ state of the CH stretching vibration in Figure~\ref{fig:Term_scheme}. It is important, however, to note that vibronic Renner-Teller coupling effects are ignored as they should not affect the  rotational and ro-vibrational spectrum for the observed $\Pi$-$\Pi$ transitions. Furthermore, coupling terms between electronic states are discarded (as is typical for such an effective Hamiltonian approach) so that potential effects of mixing between electronic states will be absorbed in the spectroscopic parameters.

The measured spectra shown below will be used to determine the molecular parameters of this model. First, the spectra will be inspected visually, in order to coarsely determine some values of the molecular parameters depicted in Figure~\ref{fig:Term_scheme}. This procedure helps in understanding the complex spectra and in assigning quantum numbers to transitions. Second, by fitting the assigned lines to the model using the PGOPHER program, parameters responsible for smaller shifts and splittings will also be determined with high confidence as discussed in a later section.

\section{Results and Discussion}
\label{sec:results}

\subsection{High-resolution IR spectrum}
\label{sec:IR}

\begin{figure*}
    \centering
    \includegraphics[width=1.0\textwidth]{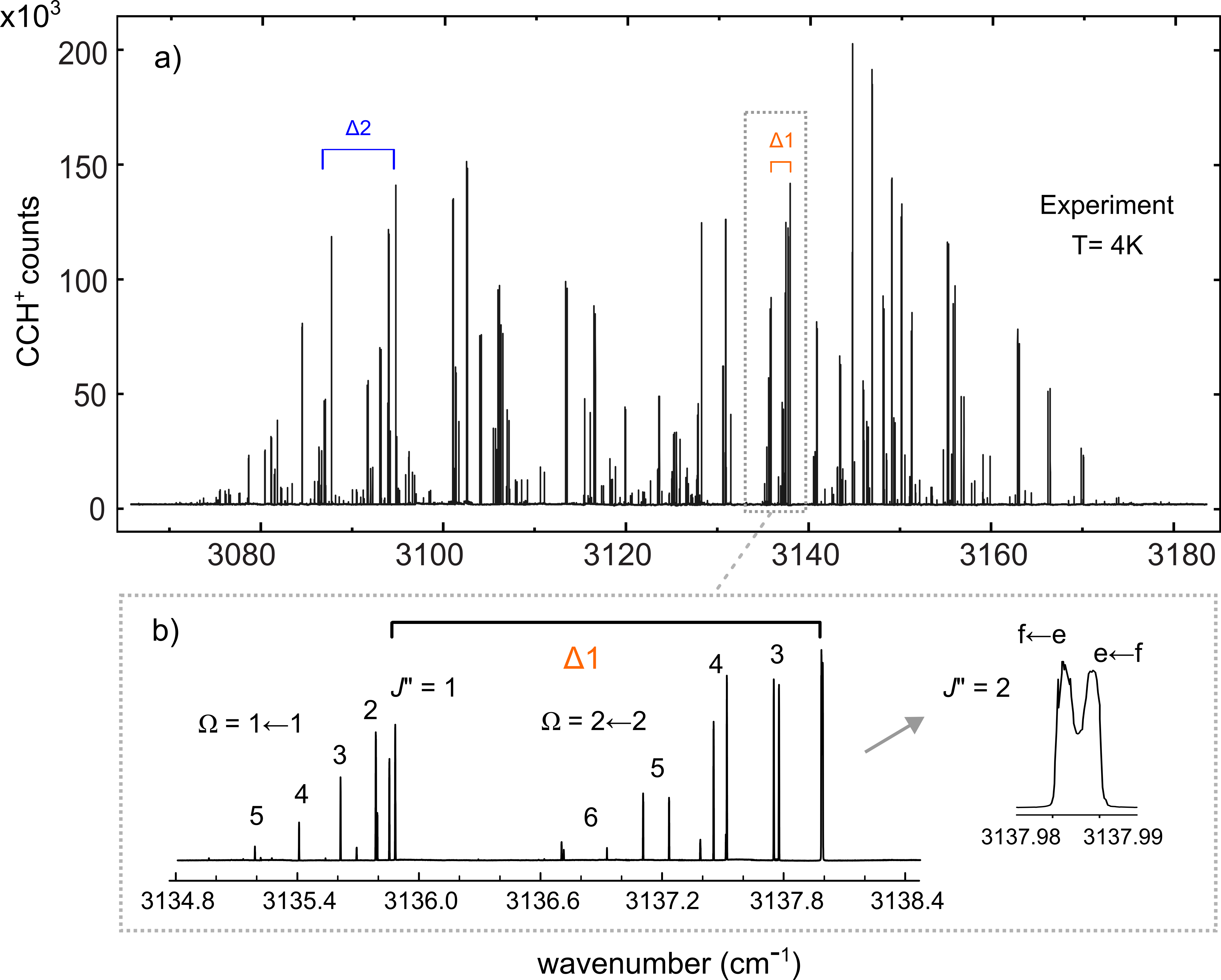}
    \caption{a) High-resolution IR spectrum of \ionp recorded by leak-out spectroscopy (LOS); \\
      b) Zoom into the 3134.8-3138.4~\wn~range showing in detail the Q-branches of the $\Omega=2$ ($F_1$) and $\Omega=1$ ($F_2$) spin-orbit components within the fundamental CH stretching vibration which are separated by roughly $\Delta1=2$~\wn. The ro-vibrational lines are nicely split into their $\Lambda$-doubling components, which for $\Omega=2$ and $J^{\prime\prime}=2$ are only partially resolved as shown in a further zoom.}
    \label{fig:LOS}
\end{figure*}

Figure \ref{fig:LOS} a) shows the high-resolution ro-vibrational spectrum of \ionp recorded by means of LOS in the range $3065.79 - 3183.59$~\wn, where the band center of the $v_1$ CH stretching vibration is expected. 
Notably, the measurement exhibits an extraordinary good signal-to-noise ratio, exceeding 2000 for the strongest feature. As a result, the spectrum shows a plethora of ro-vibrational lines, 408 in total. Each line exhibits a Gaussian line-profile which is mainly given by the Maxwell-Boltzmann distribution corresponding to the actual kinetic energy of the cold stored ions. 
Line center frequencies were obtained by fitting these Gaussian line-profiles and  are accurate to $<0.001$~\wn.
The corresponding kinetic temperature has an average value of 16.6~K, significantly higher than the nominal trap temperature of 4~K, as was also observed for other systems measured with LOS \cite{schlemmer2024high,Asvany2023,gupta23}. 
 A complete line list with signal strength is provided in the Supporting information.

No simple harmonic progression as expected for a linear molecule can be identified in the spectrum because it is the overlay of the three different fine-structure components as discussed in the previous section. 
Due to the open-shell nature of \ion, the spectrum of the $v_1$ CH stretching vibration should exhibit Q-branches, transitions with $\Delta J$ =~0. These are easy to identify when the rotational structures in the ground and excited state are similar which leads to features of a progression of close lying lines rapidly decreasing in intensity for higher $J$-values. A set of two such compact regions separated by an energy of approximately $\Delta1 = 2$~\wns are identified around 3138 \wn. This region is highlighted in Figure \ref{fig:LOS} a) and shown in Figure \ref{fig:LOS} b) as a zoom-in. There, the two progressions are clearly seen. The more intense Q-branch belongs to the lower energy $\Omega=2$ spin orbit ($F_1$) component, and the other one to $\Omega=1$ ($F_2$), which is higher in energy by the magnitude of the spin-orbit constant $|A|$. 
Due to the low temperature conditions in the trap and the overall much lower transition intensity, the third component, $\Omega=0$ ($F_3$), is not displayed in Figure \ref{fig:LOS} b). Thanks to the compactness of these Q-branches,
corresponding $J$ values can be assigned to the individual lines given in that Figure. Most of the lines come in pairs based on the $\Lambda$-doubling as will be discussed below. In fact, even for $J=2$ the $e$ and $f$ transitions are partially resolved as is shown in the right inset of Figure \ref{fig:LOS} b).

The separation of the $F_1$ and $F_2$ Q-branches is approximately $\Delta1 = 2$~\wn. Inspection of the schematic term diagram (Figure~\ref{fig:Term_scheme}) and
Eq~\ref{eqn:Hamil} reveals that the Q-branch of the $F_2$ component is approximated by the band origin of the vibrational band ($E_\mathrm{vib}$ in Eq.~\ref{eqn:Hamil})
because it does not contain a spin-orbit contribution, i.e. $S_z=0$. In contrast, the $F_1$ component ($S_z=1$) contains the difference of the SO coupling constants $A$ of the ground ($A_0$) and vibrationally excited state ($A_1$) such that for the gap between the Q-branches $\Delta1= A_1-A_0$ holds. This means that $A$ increases by about 2~\wns upon vibrational excitation. 
This value, $\Delta1$, is much smaller than the one for a second vibrational band, $\Delta2$, also being present in the observed spectrum as indicated in Figure \ref{fig:LOS} a) and discussed later. 

With the knowledge of the band origin for this band which we later associate with the $v_1$ CH stretching vibration, the corresponding P- and R-branches of the respective fine-structure (SO) components can be searched for. In fact, the values of the rotational constant $B$ in the ground and vibrationally excited $v_1$ state should not change too much. This is the reason why the lines of the Q-branches shown above are lying so close to each other. As a consequence, each SO component of the spectrum should exhibit a characteristic $2B$ structure for the different rotational levels in the corresponding P- and R-branches and more or less fixed gaps between the P-, Q- and R-branches, the value of which depends on the lowest possible $J$ value of the respective fine-structure (SO) component ($F_1, F_2, F_3$).  

\begin{figure*}[tb!]
    \centering
    \includegraphics[width=1.0\textwidth]{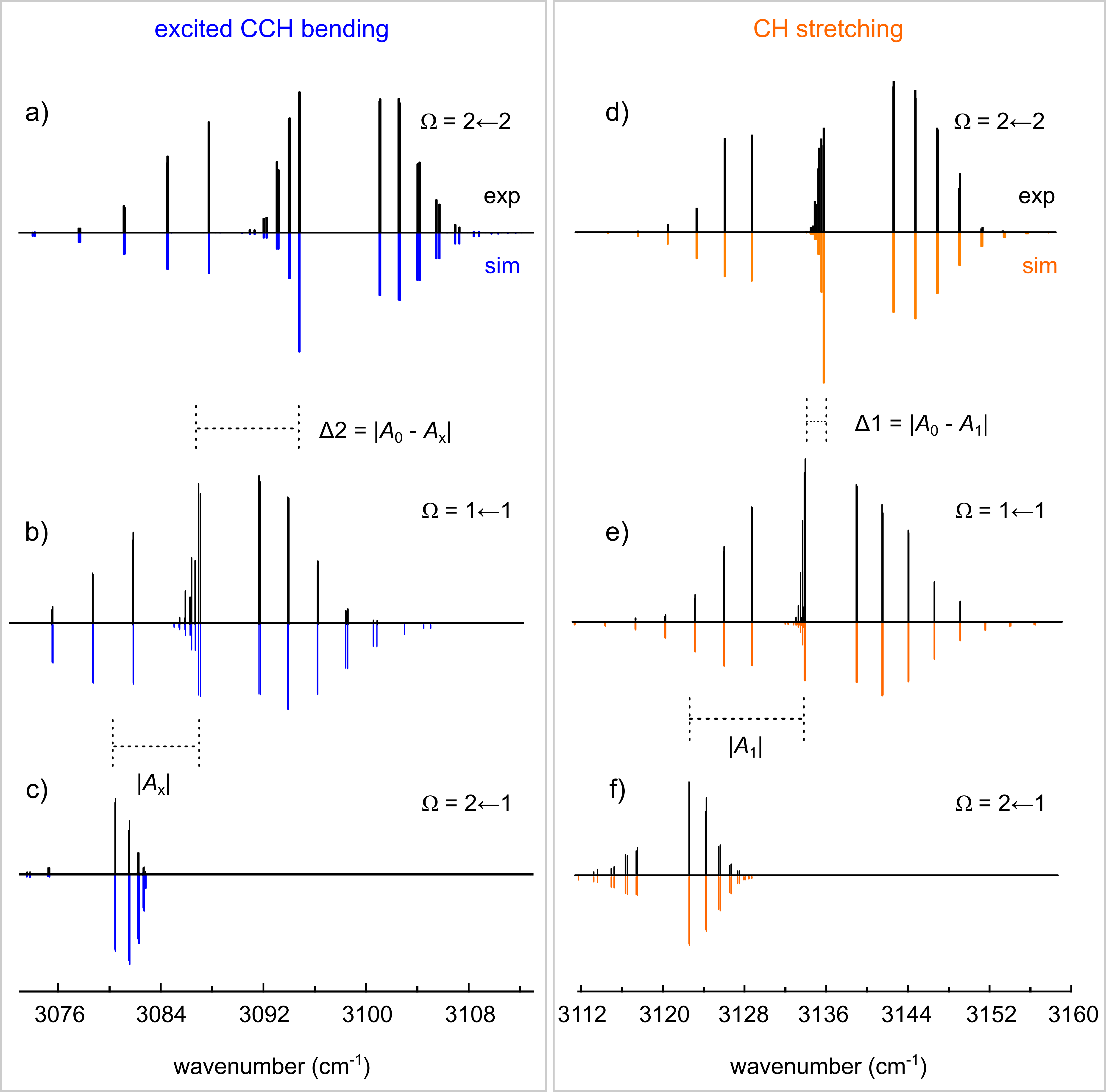}
    \caption{Rotational structure of selected spin-orbit transitions 
    of the a) - c) excited CCH bending, and d) - f) fundamental CH stretching vibration of \ion. Coarse values of the SO coupling constants can be read off from the respective band shifts as indicated by the dashed bars and as explained in the text.}
    \label{ro-vib}
\end{figure*}

This knowledge helped in identifying the sets of branches for the respective SO component. As these assignments are no longer easy to spot in 
Figure~\ref{fig:LOS} a), the identified lines  are shown in Figure~\ref{ro-vib} as a stick diagram of each line with its relative intensity. Sticks pointing upwards mark the experimentally determined line-position. Those pointing downwards belong to a simulation from a fit of the spectroscopic model to the experimental line positions.

Figure~\ref{ro-vib} a)-c) displays the rotational structure of fine-structure components for a second vibrational band in the lower frequency range ($3070 - 3125$~\wn) while Figure \ref{ro-vib} d)-f) shows these components for the band we associate with the $v_1$ vibrational band discussed so far. Figures~\ref{ro-vib} a) (b)) and d) (e)) show transitions  $F_1 -  F_1$ ($F_2 -  F_2$) where the fine structure state does not change, thus following the standard selection rule $\Delta \Omega$~=~0. 
Again, as already shown in  Figure \ref{fig:LOS} the band shifts $\Delta2$ and $\Delta1$ discussed above appear also in this Figure as indicated by the horizontal bars between Figure\ref{ro-vib} a) and b) as well as d) and e), respectively. 

Indeed, the spectra displayed in Figure \ref{ro-vib} d) and e) show the $2B$ spacings in the P- and R-branches corresponding to a rotational constant of $B~\approx$ 1.35~\wns which is in line with the theoretical prediction by 
\citet{Mehnen2018} (1.3669~\wn, corresponding to 40.980 GHz). In fact, the good agreement between the experimental stick spectrum and the simulation shows that not only the rotational constants but also distortion constants and other molecular constants can be determined with high precision. 

However, those spectra also show that this band only extends from around 3110~\wns to 3160~\wn. Therefore, all lines below 3110~\wns belong apparently to a different vibrational band. As discussed above, this other band centers around 3090~\wns and also exhibits Q-branches, whose lines however are not lying as close as for the $v_1$ band. 
The gap between the Q-branches is approximately $\Delta2 = 8$~\wn, as shown in Figure~\ref{fig:LOS} a) as well as in Figure~\ref{ro-vib} a) and b), saying that the SO constants $A$ differ much more for this band, $\Delta2= A_x-A_0$. Also, the spread of lines in the Q-branches is larger, showing that the change in the rotational constants from the ground to the excited state is considerably larger for this band than for the $v_1$ band discussed so far. 

The structure of this unknown band has been fitted to the same $^3\Pi-^3\Pi$ rovibronic effective Hamiltonian model thus leading to a list of corresponding molecular constants. Both bands have the same ground state combination differences which shows that they originate from the same ground state, $v=0$, which is expected in a low temperature experiment as presented here. Therefore, fitting the whole spectrum with these two bands led to three sets of constants for the ground and the two vibrationally excited states which are listed in Table~\ref{tab:Fitted_param},  as will be discussed in Section \ref{sec:fit_param} in detail. 
Following the above described procedure the main parts of the spectrum could be assigned and fitted. 

As it turns out, transitions changing the fine structure state, e.g., $F_1 -  F_2$, are allowed for \ionp and are shown in Figures~\ref{ro-vib} c) and f) for the two vibrations discussed here.  An overview of these $\Delta \Omega\neq0$ transitions is shown in the Supporting Information, Figure~S7. Clearly, the calculated intensities are (extremely) under-predicted in the employed effective Hamiltonian model, where electronic state coupling terms are ignored. The $\Delta\Omega\neq0$, or cross-$\Omega$, transitions likely gain intensity through mixing with the low-lying $^3\Sigma^-$ state. This aspect is thoroughly discussed in an accompanying paper\cite{Steenbakkers2025}.

 Here, we just fit the additional branches to our model which also accommodates so-called cross-$\Omega$ transitions. Again, inspection of the schematic term diagram, Figure~\ref{fig:Term_scheme}, reveals that the  $F_2$ ($\Omega=1$) -- $F_1$ ($\Omega=2$) transitions in Figures~\ref{ro-vib} c) and f) only contain the $A$ constant of the $F_1$ ($\Omega=2$) component while it does not appear for $F_2$ ($\Omega=1$). Therefore, when comparing the Q-branches  of the  $F_2 -  F_2$ transitions in Figures~\ref{ro-vib}~b) and e) to the respective ones for the $F_2-F_1$ transitions in Figures~\ref{ro-vib}~c) and f), they should be shifted by the SO constant $A$ of the vibrationally excited state as is indicated by a horizontal bar between Figures~\ref{ro-vib}~b) and c), as well as e) and f). Note that here the Q-branch transitions of panel b), resp. e), are compared to the R-branch transitions of panel c), resp. f), since $J$ is increasing when $\Omega$ increases while the molecular rotation remains unchanged. As a result the shifts for the two bands amount to $A \approx -12$ and $-6$~\wns for the $v_1$ band and the unknown  other band, respectively. Together with the differences of $A$ inferred from $\Delta1$ and $\Delta2$ the ground state value of 
the SO coupling constant is close to $A=-14$~\wn.

While this "pedestrian" way of assigning the spectrum is cumbersome it clearly shows that the main spectroscopic parameters can be inferred directly from the spectrum  and their values are determined to high accuracy from a combined fit of the two bands discussed so far (Section \ref{sec:fit_param}). Moreover, the differences in $A$ aid the assignment of the two vibrational bands shown in Figure~\ref{fig:LOS}.
For completeness, all of the isolated cross-$\Omega$ progressions are shown in the supporting information in Figures S9 -- S18 together with their calculated line positions.

After incorporating all cross-$\Omega$ terms, only 18 out of the 408 lines shown in Figure~\ref{fig:LOS} remain unassigned, with normalised intensity ranging from 0.1~\% to 4~\% of the strongest feature. These unidentified lines most probably correspond to lines from nearby vibrational modes\cite{Steenbakkers2025}, such as bending overtones, or could be a result of some missing resonances or interactions in the used model. Since these (few) lines are all (very) weak and do not change the overall interpretation of the observed spectrum, no further investigation was deemed relevant.

Before a more thorough discussion of the derived spectroscopic parameters, the pure rotational spectra are presented in the next section.

\subsection{Rotational measurements}
\label{sec:Rot_meas}

\begin{figure*} [h]
    \centering
    \includegraphics[width=0.85\textwidth]{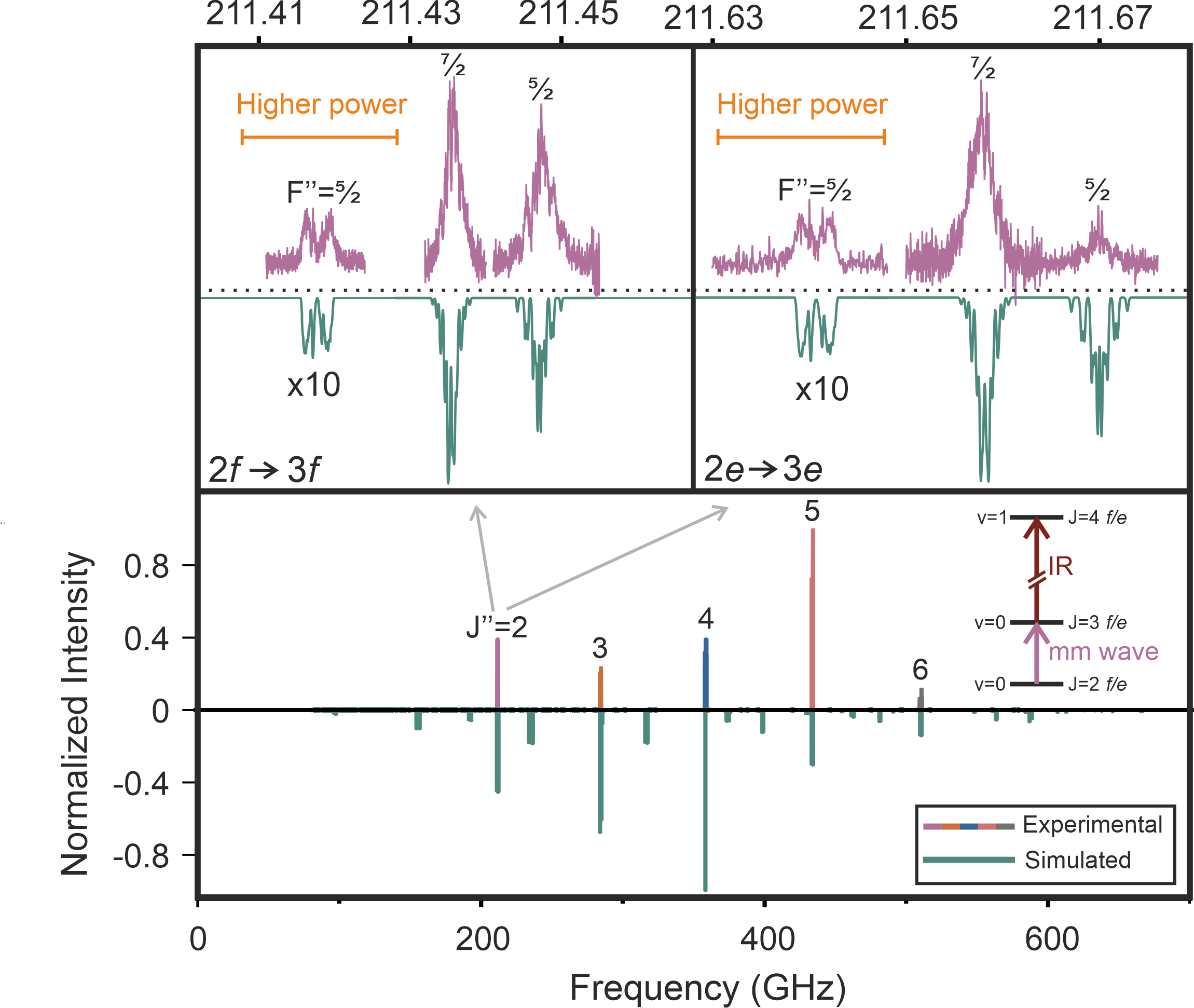}
    \caption{(lower panel) Overview of the measured pure rotational transitions, with the different colours representing a single $J^{\prime\prime}$ (also given in Table \ref{tab:Rot-lines}), together with the simulated spectrum (negative axis, 20~K), based on the spectroscopic constants obtained from a global fit including the ro-vibrational and rotational transitions. 
    (upper panels) Zoom of the measured $J = 3\leftarrow2$ transition revealing $\Lambda$-doubling, nuclear hyperfine structure as well as Zeeman splitting. Similar splittings were found for the other transitions (see Supporting Information Figures S19 -- S22).}
    \label{fig:rot_lines}
\end{figure*}

Based on the ground state spectroscopic parameters obtained from a preliminary fit of the ro-vibrational data, we searched for pure rotational transitions of \ionp using double-resonance vibrational-rotational spectroscopy based on LOS. This method has been very successful in recording rotational transitions of other ions of astrochemical interest\cite{sil24,sil23,gupta23,Asvany2023,sil25}. For \ion, a total of 5 rotational transitions (from $J^{\prime\prime} = 2$ to $J^{\prime\prime} = 6$), each with resolved $\Lambda$-doubling and hyperfine splitting components, within the lowest fine-structure state, $\Omega = 2$, were measured in the range $200-520$~GHz. A permanent dipole moment of 1.06 D for the vibrational ground state was calculated at the MRCI/ANO1 and HF/cc-pVQZ level of theory. An overview of the observed transitions is provided in Figure \ref{fig:rot_lines} and Table \ref{tab:Rot-lines}. 

While recording the transitions shown in Figure \ref{fig:rot_lines}, the wavenumber of the IR laser was kept fixed on resonance with a ro-vibrational transition which starts from a specific rotational level in the ground vibrational state (red-brown arrow in the energy level scheme of Figure \ref{fig:rot_lines}). Then, the mm-wave radiation (pink arrow in the level scheme of Figure~\ref{fig:rot_lines}) was used to excite a pure rotational transition within $v = 0$ that ends in this common rotational level probed by the IR laser. This results in an increase in the population of this common level and consequently, an increase in the LOS signal is observed. Each transition was recorded in several individual measurements in which the mm-wave frequency was tuned in a given window while monitoring the ion counts. The step size for the frequency scans was typically 50 or 100~kHz when searching for lines and 20 kHz for final measurements. The rotational spectrum was then normalized by dividing the number of \ionp ions in the scanned frequency window by those at an off-resonance position. Thus, the baseline of the observed features is close to unity and is subtracted from the experimental data. The intensity of the most intense rotational transition ($J = 6\leftarrow5$) is set to 1 with all other intensities given relative to this value, and the resulting averaged spectra are shown in Figure \ref{fig:rot_lines}. To minimize power broadening effects, the power of the mm-wave radiation was optimized for each rotational transition and lowered as much as possible. For the $J = 3\leftarrow2$ transition for example, higher power was needed to observe its less intense components, while the power was attenuated for the most intense features, as indicated in the inset of Figure ~\ref{fig:rot_lines}. One can therefore not directly compare the measured relative intensities to the simulated ones.

\begin{table*}[htb!]
\caption{\label{tab:Rot-lines} Observed rotational transitions of \ionp with the experimental uncertainty of the transition frequencies given in parentheses.}
\begin{threeparttable}

    \begin{tabular}{l r c c c c c c c}
    \hline
Frequency  &  o-c & $J^\prime$  & $F^\prime$ & $p^\prime$ & $J^{\prime\prime}$ & $F^{\prime\prime}$ & $p^{\prime\prime}$ & IR line \\
 (MHz) & (kHz) & & & & & & & (\wn)\\
\hline\hline
211417.82(8)  & 60 &  3   & 2.5&  f & 2   & 2.5 &  f  & 3146.951(1) \\ 
211435.49(8)  & 41  &  3   & 3.5&  f & 2   & 2.5 &  f  & 3146.951(1) \\ 
211447.14(8)  & 23  &  3   & 2.5&  f & 2   & 1.5 &  f  & 3146.951(1) \\ 
211640.65(8)  & $-$82 &  3   & 2.5&  e & 2   & 2.5 &  e  & 3146.981(1) \\ 
211658.09(8)  &  $-$44 &  3   & 3.5&  e & 2   & 2.5 &  e  & 3146.981(1) \\ 
211670.03(8)  & 34  &  3   & 2.5&  e & 2   & 1.5 &  e  & 3146.981(1) \\ 
283991.43(8)  &  10 &  4   & 3.5&  f & 3   & 3.5 &  f  & 3149.094(1) \\ 
284002.47(8)  &  $-$28 &  4   & 4.5&  f & 3   & 3.5 &  f  & 3149.094(1) \\ 
284009.07(8)  &  $-$39 &  4   & 3.5&  f & 3   & 2.5 &  f  & 3149.094(1) \\ 
284449.30(8)  & 40   &  4   & 3.5&  e & 3   & 3.5 &  e  & 3149.147(1) \\ 
284459.78(8)  & 18  &  4   & 4.5&  e & 3   & 3.5 &  e  & 3149.147(1) \\ 
284466.66(8)  & $-$2  &  4   & 3.5&  e & 3   & 2.5 &  e  & 3149.147(1) \\ 
357927.03(8)  &  $-$6  &  5   & 4.5&  f & 4   & 4.5 &  f  & 3151.246(1) \\ 
357933.82(8)  &  $-$4  &  5   & 5.5&  f & 4   & 4.5 &  f  & 3151.246(1) \\ 
357938.07(8)  & $-$44  &  5  & 4.5&  f & 4   & 3.5 &  f & 3151.246(1) \\
358660.12(8)  &  $-$3  &  5  & 4.5&  e & 4   & 4.5 &  e & 3151.326(1) \\
358665.98(8)  &   0    &  5  & 5.5&  e & 4   & 4.5 &  e & 3151.326(1) \\
358670.60(8)  & $-$25  &  5  & 4.5&  e & 4   & 3.5 &  e & 3151.326(1) \\
433167.16(8)  & 3     &  6  & 5.5&  f & 5   & 5.5 &  f & 3153.416(1) \\
433171.02(8)  & 55    &  6  & 6.5&  f & 5   & 5.5 &  f & 3153.416(1) \\
433173.94(8)  & $-$5  &  6  & 5.5&  f & 5   & 4.5 &  f & 3153.416(1) \\
434182.58(8)  & $-$15 &  6  & 5.5&  e & 5   & 5.5 &  e & 3153.532(1) \\
434185.10(8)  &  16 &  6  & 6.5&  e & 5   & 5.5 &  e & 3153.532(1) \\
434188.48(8)  &  27 &  6  & 5.5&  e & 5   & 4.5 &  e & 3153.532(1) \\
509555.19(12) &  16    &  7  & 7.5&  f & 6   & 6.5 &  f & 3155.729(1) \\
509557.29(12) & $-$41  &  7  & 6.5&  f & 6   & 5.5 &  f & 3155.729(1) \\
510838.87(12) & $-$31 &  7  & 7.5&  e & 6   & 6.5 &  e & 3155.504(1) \\
510841.46(12) &  7    &  7  & 6.5&  e & 6   & 5.5 &  e & 3155.504(1) \\
\hline
    \end{tabular}
    \end{threeparttable}
\end{table*}

It is worth reinforcing that all measured rotational transitions within the lowest energy $\Omega = 2$ state ($F_1$) showed well resolved $\Lambda$-doubling components ($e$ and $f$) each displaying hyperfine splittings due to nuclear spin-orbit and nuclear spin -- electron spin coupling as well as nuclear spin -- electron spin dipole-dipole interactions involving the H ($I = 1/2$) nucleus (see Figure \ref{fig:Term_scheme} for an energy level diagram). Apart from $\Lambda$-doubling and hyperfine splitting, additional substructure was observed in the spectra which was attributed to Zeeman splitting likely caused by the magnetically levitated pumps used in our experimental setup\cite{Marimuthu2022,sil25}.
Overall, the Zeeman splitting appeared as a substructure contributing to a substantial broadening of all observed transitions, especially for the high $J$ transitions ($6\leftarrow5$ and $7\leftarrow6$), where the Zeeman splitting could not be resolved. 
The structure and shape of each of the lines allow us to estimate the magnetic field as well as its relative orientation, which could be determined as a ratio between parallel and perpendicular polarization. 
This was done using the built-in functionality of PGOPHER to simulate the Zeeman pattern and its value was adjusted by eye until it reproduced the observed spectral patterns (individually for each transition).
Though it seems like the magnetic field varies slightly between measurements, the overall strength of the magnetic field was estimated to be about 0.00018(2)~T, which is roughly 3.5 times the Earth's magnetic field and the polarization ratio to be about 1/1.6 for parallel/perpendicular orientation. The experimental results together with the simulated spectra are shown in Figure~\ref{fig:rot_lines} for the $J = 3\leftarrow2$ transition and in the supporting information Figures S19 -- S22 for the other transitions. To obtain the frequencies of the unsplit lines the simulated Zeeman pattern was manually overlayed with the experimental spectrum, after which the magnetic field of the simulation was set to zero. The error of this procedure was estimated to be about 80 kHz for all transition frequencies except for the $J = 7\leftarrow6$ transition, which was given an error of 120~kHz, due to the relatively noisy appearance of this transition. The such derived center frequencies of all rotational lines are given in Table \ref{tab:Rot-lines}. 

It is worth mentioning that apart from allowing pure rotational transitions of \ionp to be recorded, the double-resonance spectroscopy scheme has the ability to determine the connectivity of transitions and thus to confirm the ro-vibrational assignments \cite{sil23b}. Here, it was crucial to confirm the ro-vibrational assignments for the $R$-branch $v = 1$ $J^\prime = 8$ lines of the CH stretching mode within the $\Omega = 2$ state, which turned out to be perturbed, causing a reversal of the ordering of the parity ($e$ and $f$) of the $\Lambda$-doubling components. This was verified by performing double-resonance experiments in which the search for the $7\leftarrow6$ transitions (both $e$ and $f$ components) at around 510~GHz was performed by keeping the IR laser fixed at each of the candidate IR transitions at 3155.729(1) ($f$) and 3155.504(1) \wns ($e$) at a time. Since a rotational transition will only be observed when the correct ro-vibrational line connecting the common probed state in the ground state is used, the IR assignments could be confirmed.

\subsection{Spectroscopic analysis}
\label{sec:fit_param}

From the 390 assigned IR ro-vibrational lines 385 were included 
in the fit (3 tentative lines and the 2 distorted $R$-branch $v = 1$ $J^\prime = 8$ lines were excluded), 
along with the ground-state rotational transitions from $J^{\prime\prime} = 2$ to $J^{\prime\prime} = 6$. 
For the IR lines an uncertainty of 0.001~\wns (30 MHz) was assumed, 
which is primarily determined by the accuracy of the wavemeter. 
This uncertainty is considerably larger than that of the rotational lines (80 or 120 kHz). 
Moreover, the linewidth of the ro-vibrational lines is dominated by the Doppler broadening, 
with an average full-width half-maximum of approximately 52~MHz. 
This linewidth allows the $\Lambda$-doubling to be resolved but  prevents 
resolving the hyperfine structure caused by nuclear spin (<30 MHz). 
To unify the hyperfine-resolved rotational features 
and the ro-vibrational lines in a global fit,
the 'removespins' option in PGOPHER~\cite{wes17} was used.

The fitting process involved iteratively adding and removing ro-vibrational 
transitions from the linelist while closely monitoring the total weighted 
error and the standard deviation of the spectroscopic parameters. 
If a particular feature or set of features significantly degraded the quality of the fit, most probably due to perturbations in the upper state, 
they were excluded. Once the dataset was finalized, 
a ground-state combination difference (GSCD) analysis was performed 
to determine thirteen vibrational ground-state parameters,  presented 
in the second column of Table~\ref{tab:Fitted_param}.
The weighted average error of the GSCD fit was 0.48, 
indicating that (i) the Hamiltonian used describes the 
experimental spectrum rather well, and that (ii) our experimental uncertainties (rotational as well as ro-vibrational) are somewhat overestimated. The terms describing the hyperfine interaction are the  nuclear spin-orbit and nuclear spin -- electron spin coupling  parameters $a$ and $b$, respectively, while dipole-dipole interactions between the nuclear and electronic spins were given by parameters $c$ and $d$. The $c$ parameter, however, did not converge with the number of recorded lines and was set to zero. For the other parameters, the values obtained were $a=50.363(95)$~MHz, $b=-11.504(52)$~MHz, and $d=-7.22(28)$~MHz.

\begin{table*}[!tb]
\caption{\label{tab:Fitted_param} Spectroscopic constants for \ionp from a global fit of the ro-vibrational and rotational spectrum together with the standard deviation in parentheses. All parameters are given in MHz unless otherwise specified.}
\begin{threeparttable}

    \begin{tabular}{l  S[table-format=7.8] S[table-format=7.8]  S[table-format=7.8]}
    \hline\hline
      Parameter & {$v = 0$} & {$v_1 = 1$} & {excited CCH bending\cite{Steenbakkers2025} } \\
     \hline
     $\nu$ (\wn) & 0            &  3136.12203(18) & 3087.59545(18) \\
     $A$ (\wn)   & -13.831361(69) & -11.87874(37)  & -5.95785(38) \\
     $A_D$       & 4.45(15)     & -36.8(26)       & 189.9(22) \\
     $A_H$       & {--}         & -1.860(91)      & -17.39(16) \\
     $B$         & 40649.48(11) & 40019.3(1.5)    & 39254.97(93)   \\
     $D$         & 0.10715(35)  & -0.377(22)      & -0.334(18) \\
     $o$         & -17129.3(2.1) &  -15799.3(7.6) & -10518.0(7.9) \\
     $o_D$       & {--}         & -5.07(42)       & 2.40(42)\\
     $p$         & 64.40(49)    & 85.7(2.7)       & -24.9(3.1) \\
     $p_D$       & 0.0745(37)   & 0.651(96)       & 3.30(11) \\
     $q$         & -95.176(41)  & -152.21(53)     & -271.20(87) \\
     $q_D$       & {--}         &  {--}           & -0.532(23)  \\
     $\lambda_{SS}$ & -4451.9(1.2) & -1141.9(7.3) & -7768.0(6.1) \\
     $\lambda_D$ & -0.365(74)   & -11.3(1.1)      & {--} \\
     $\gamma_D$  & {--}         & 9.03(81)        & 48.38(50) \\
     $\gamma_H$  & {--}         &  0.0499(82)     & 0.2200(49)  \\
     $a$         & 50.363(95)   &  {--}           & {--}  \\
     $b$         & -11.504(52)  &  {--}           & {--}  \\
     $c$         & {--}         &  {--}           & {--}  \\
     $d$         & -7.22(28)    &  {--}           & {--}  \\
    \hline\hline
    \end{tabular}
    \end{threeparttable}
\end{table*}

The parameters for the excited vibrational states were then obtained by fixing the 
ground-state parameters and fitting only the excited-state parameters, 
resulting in a weighted error of 0.96. 
Those final spectroscopic parameters, thirty in total,  are presented in 
the last two columns of Table~\ref{tab:Fitted_param}.
For reference, the total weighted error when including both 
ground- and excited-state parameters in the fit was 0.97 
(with 43 parameters fitted in total). The higher weighted errors when including the upper states (0.96 compared to 0.48) suggests that the upper levels are slightly perturbed, see also the discussion on spectroscopic parameters later on.

With respect to the fitted parameters, the lines belong to two vibrational modes with fitted band centers 3087.59545(18) and 3136.12203(18)~\wn. Comparing these values to earlier theoretical and experimental works (summarized in Table~\ref{tab:Frequencies}), it is likely that the band at 3136.12203(18)~\wns corresponds to the CH stretching mode ($\nu_1$). It is important to note that all of the calculated frequencies are purely in harmonic approximation, which explains their blue shift compared to the experimental values. Furthermore, the frequencies obtained from the He-droplet \cite{Feinberg2023} and Ar-matrix isolation \cite{Andrews1999} experiments may be shifted from our values due to interactions between the \ionp and the used solvation medium. 

The band at 3087.59545(18)~\wn, which was likely observed previously by \citet{Feinberg2023} at 3111(1) \wn, is more challenging to assign. Based on the spectroscopic fit described above we can deduce that this feature is of $\Pi$ symmetry, suggesting that it could correspond to the overtone of the CC stretch ($\nu_3$). However, based on the three-state diabatic model described by \citet{Steenbakkers2025}, it was (tentatively) assigned to a highly excited and mixed overtone of a Renner-Teller and Pseudo-Jahn-Teller split CCH bending mode. Despite the slight uncertainty in assigning this band, its individual ro-vibrational components were assigned and included in the spectroscopic model, leading to a more precise characterization of the ground-state spectroscopic parameters.

\begin{table*}[!tb]

\caption{\label{tab:Frequencies}Overview of calculated and measured vibrational frequencies of \ionp (in \wn). Here $\nu_1$ represents the CH stretch, $\nu_2$ the CCH bending and $\nu_3$ the CC stretch.}
    \begin{threeparttable}

    \resizebox{\textwidth}{!}{\begin{tabular}{l c c c c c}
    \hline\hline
      Method & $\nu_1$ & $\nu_2$ & $\nu_3$ & other &ref \\
     \hline
     LOS & 3136.1220(2) & ... & ... & 3087.5954(2)& This work\\
     He-droplet & 3145(1) & ... & ... & 3111(1)/3183(1) & expt.\cite{Feinberg2023}\\
     SPES& ... & ... & 1620(40) & ... & expt.\cite{Gans2017}\\
     Ar-matrix isolation & ... & ... & 1832.2(5) & ... & expt.\cite{Andrews1999}\\
     PBE0/aug-cc-pVDZ    & 3274 &  447/755 & 1914 & ... &  calc.\cite{Mehnen2018}\\
     RCCSD(T,full)/cc-pwCVTZ    & 3274  &  415/802 & 1849 & ... &  calc.\cite{Mehnen2018}\\
     RCCSD(T)-F12/cc-pVQZ-F12   &  3272  &  438/801 & 1853 & ... &  calc.\cite{Mehnen2018}\\
     CASSCF/aug-cc-pVQZ & 3197 & 446/805 & 1808 & ... & calc.\cite{Gans2017}\\
     MRCI/aug-cc-pVQZ & 3242 & 347/719 & 1839 & ... & calc.\cite{Gans2017}\\
     CCSD(T)/aug-cc-pVDZ & 3247 & 747/201 & 1831 & ... & calc.\cite{Andrews1999}\\
     BP86/6-31+G** & 3207 & 745/129 & 1834 & ... & calc.\cite{Andrews1999}\\
     B3LYP/6-31+G** & 3284 & 782/435 & 1902 & ... & calc.\cite{Andrews1999}\\
     \hline\hline

    \end{tabular}}
    \end{threeparttable}
\end{table*}

From the other spectroscopic parameters, the rotational constant $B$ and the spin-orbit (SO) constant $A$ are the only parameters for which prior data exists. The fitted rotational constant of the vibrational ground state ($B=40.64948(11)$ GHz corresponding to 1.3559~\wn) aligns relatively well with values derived from the geometries reported by \citet{Mehnen2018}: 40.609~GHz [RCCSD(T)/cc-pVTZ-DK], 41.044 GHz [RCCSD(T)-F12/cc-pVQZ-F12], and 40.980 GHz [RCCSD(T, full)/cc-pwCVTZ]. Regarding the SO constant, the previous experimental value of $|A| = 15$\wn\ reported by \citet{Gans2017}, obtained via slow photoelectron spectroscopy (SPES), as well as the calculated value of $A = -18$\wn\ from \citet{Mehnen2018}, computed at the MRCI/aug-cc-pVTZ level of theory, are in reasonable agreement with our fitted ground state value, $A = -13.831361(69) $~\wn. 
 
 Significant changes of the SO constant are observed upon vibrational excitation. 
 For the band we assign to the $v_1$ CH stretching vibration $A$ changes by 16~\%, but by 57~\% for the band associated with the CCH bending vibration. In fact, a bending vibration is expected to have a stronger influence on the SO interaction and therefore, this change in $A$ is taken as additional aid for the assignment of the two vibrational bands studied here. Overall, the significant changes observed in the SO constant upon vibrational excitation are likely attributed to mixing with the low-lying $^3\Sigma^-$ state, resulting in the reduction of the $\Pi$-character of the vibrational level. This phenomenon is described in detail in an accompanying paper\cite{Steenbakkers2025}. We should also note that the rotational constant for $v_1$ is changing by only $\sim$1.5~\% while it changes by 3.6~\% for the other band. 
 
The present spectroscopic analysis reproduces the experimentally observed transition frequencies with an accuracy well within their experimental uncertainty. However, it is important to note that high-resolution rotational transitions were only observed in the vibrational ground state of the lowest $\Omega = 2$ fine-structure state. While predictions for higher $J$ rotational transitions within this state should remain reliable, uncertainties may arise in predictions involving transitions within or between other fine-structure states due to missing terms in the effective Hamiltonian and less accurate spectroscopic constants in these states.

It is also essential to emphasize that the model employed in this analysis is empirical in nature. Although the model effectively simulates the observed spectrum, the physical interpretation of the derived spectroscopic parameters must be approached with some caution. This is primarily due to the exclusion of the low-lying $^3\Sigma^-$ state from the effective Hamiltonian, which could have a considerable impact on several interactions, particularly spin-orbit coupling, $\Lambda$-doubling, and spin-spin coupling. The parameters used in this model may absorb some of these unaccounted effects, complicating their interpretation. For example, the negative sign of the centrifugal distortion constants for the excited vibrational states, and the inability to successfully include higher-order centrifugal distortion constants, suggest that this parameter might be absorbing interactions not explicitly considered in the model. To extract physically meaningful interpretations of the parameters, it would be necessary to explicitly incorporate the $^3\Sigma^-$ state into the Hamiltonian, a task that lies beyond the scope of this work.

\section{Conclusions}

In this work we have presented the first high-resolution IR spectrum of the open-shell linear ion \ionp in the range $3065.79-3183.59$~\wn, covering the CH stretch fundamental and an additional band exhibiting a $\Pi$-$\Pi$ transition. This feature was tentatively assigned to a highly excited bending overtone using a three-state diabatic model that includes the two ground $^3\Pi$-states and the low-lying $^3\Sigma^-$ state, which is discussed in detail in another publication\cite{Steenbakkers2025}. 

The richness of the ro-vibrational spectrum shown in Figure~\ref{fig:LOS} and the absence of a simple rotational structure illustrates the complexity of spectra for open-shell molecules, even for a linear molecule like \ion. In fact, this work has been started about ten years ago in the Cologne laboratories using the methods of laser induced inhibition of cluster growth (LIICG, \cite{Chakraborty2013,Asvany2014}) and rotational state dependent attachment of rare gas atoms (ROSAA, \cite{Brunken2014,Brunken2017}) to record the pure rotational spectrum. Thanks to the advent of LOS the signal-to-noise ratio was improved to the level shown in Figure~\ref{fig:LOS} unfolding also many more and weaker features.  

The complete spectrum comprises 408 lines, a number still limited thanks to the low temperature environment of the trap experiment which helped the final assignment. We successfully assigned 390 lines, many of which could be attributed to several $\Delta\Omega \neq 0$ transitions, typically forbidden for $\Pi$-$\Pi$ transitions, whose presence are likely a result of electronic state mixing.

The resulting spectroscopic model aligns very well with the experimental data, providing a reliable and accurate set of spectroscopic constants. LOS was also instrumental to study the spectra of other fundamental open shell ions including HCN$^+$ ($^2\Pi$) and  HNC$^+$ ($^2\Sigma$) \cite{scm25}, for which also pure rotational spectra \cite{sil25} and rotationally resolved electronic spectra \cite{mar25,Redondo2026} were recorded and analysed. Therefore, it seems quite feasible to also investigate the low-lying $^3\Sigma^- \leftarrow X^3\Pi$ electronic transitions of \ionp with rotational resolution. 

The ground-state constants of \ionp derived from the IR fit facilitated the first search for the pure rotational spectrum of \ion. We observed five rotational transitions ranging from $J^{\prime\prime}=2-6$ within the $\Omega=2$ ground state. Based on these measurements, \ionp has been detected in space for the first time \cite{Jacob2025}. The observation of the pure rotational spectrum not only provides detailed information on the resolved hyperfine structure of \ion, but also confirms the validity of the employed spectroscopic model. Collectively, the ro-vibrational and pure rotational spectra, along with the spectroscopic model presented here, offer a comprehensive spectroscopic description of \ion. 
The current data set will facilitate future astronomical detections, whether by radio-astronomy similar to the detections of \emph{l}-C$_3$H$^+$\cite{pety2012iram,Brunken2014}, HC$_3$S$^+$ \cite{Cernicharo2021} and H$_2$C$_3$H$^+$ \cite{sil23,sil23b}, or via infrared astronomy, as recently demonstrated for CH$_3$$^+$\cite{berne2023,changala2023}.

\section*{Supporting Information}

Supporting information files include: additional ground-state rotational as well as ro-vibrational spectra for different $\Omega$-progressions within both the CH fundamental and excited CCH bending overtone of \ionp (PDF); PGOPHER simulations with corresponding spectroscopic fits for the different vibrational states (ZIP); measured ro-vibrational spectra in the $\approx3065-3183$~\wn range (ZIP).

\section*{Acknowledgements}
This paper is dedicated to the memory of our colleague and friend John F. Stanton. His profound understanding of molecular spectroscopy and unwavering optimism and enthusiasm will be deeply missed. This work has been started about ten years ago using the methods of laser induced inhibition of cluster growth (LIICG) and rotational state dependent attachment of rare gas atoms (ROSAA), supported by the "Cologne Center for Terahertz Spectroscopy" (DFG SCHL 341/15-1) and via the priority program SPP 1573 (DFG BR 4287/1-2). In recent years it has been part of the research program “HFML-FELIX: a Dutch Center of Excellence for Science under Extreme Conditions” (with Project No. 184.035.011) of the research program “Nationale Roadmap Grootschalige Wetenschappelijke Infrastructuur,” which is partly financed by the Netherlands Organisation for Scientific Research (NWO). This work has been supported by an ERC Advanced Grant (MissIons: 101020583), and by the Deutsche Forschungsgemeinschaft (DFG) via the Collaborative Research Centre 1601 (project ID: 500700252, sub-projects C4 and B8). The Toptica cw-OPOs have been financed by HBFG (INST216/1184-1, project number 504504934 and INST216/1069-1, project number 450096019). WGDPS thanks core funding from the University of
Cologne and the Alexander von Humboldt Foundation for funding through a Postdoctoral Fellowship during the time this work has been carried out.

\bibliography{references,coltrap}

\newpage
\begin{center}
    Table of Contents Graphic
\end{center}

\begin{figure} [h]
    \centering
    \includegraphics[width=0.7\textwidth]{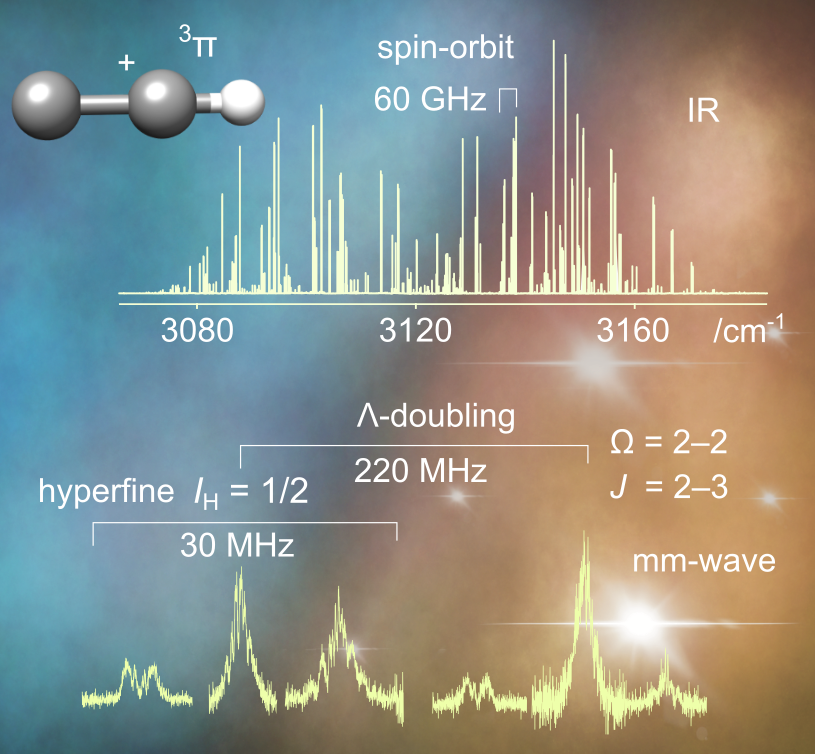}
\end{figure}

\end{document}